%% file: main.tex
\documentclass[10pt,twocolumn,letterpaper]{article}

\usepackage[ruled]{algorithm2e} 

\usepackage[accsupp]{axessibility} 

\usepackage{cvpr}              
\usepackage{graphicx}
\usepackage{booktabs}
\usepackage{amssymb}
\usepackage{soul}
   
\usepackage{amsmath, amsfonts, amsthm, bm}  
\usepackage{cuted}
\usepackage{capt-of}
\usepackage{multirow}
\usepackage{xcolor}
\usepackage{upgreek}

\usepackage{makecell}
\usepackage{amstext}
\usepackage{algpseudocode}
\usepackage{bbding}
\usepackage{lipsum}


\definecolor{cvprblue}{rgb}{0.21,0.49,0.74}
\usepackage[pagebackref,breaklinks,colorlinks,allcolors=cvprblue]{hyperref}

\def\paperID{9779} 
\def\confName{CVPR}
\def\confYear{2025}

\title{Creating Your Editable 3D Photorealistic Avatar with\\ Tetrahedron-constrained Gaussian Splatting}

\author{Hanxi Liu$^1$, Yifang Men$^2$, Zhouhui Lian$^1$\thanks{Corresponding author.  E-mail: lianzhouhui@pku.edu.cn\\This work was supported by the National Natural Science Foundation of China (Grant No.: 62372015), Center For Chinese Font Design and Research, and Key Laboratory of Intelligent Press Media Technology.}
\\
$^1$Wangxuan Institute of Computer Technology, Peking University, China\\
$^2$Institute for Intelligent Computing, Alibaba Group\\
}

\begin{document}
\maketitle

\begin{strip}
    \centering
    \vspace{-2.2cm}
    \setlength{\abovecaptionskip}{-0.1cm}
    \includegraphics[width=1.0\textwidth]
{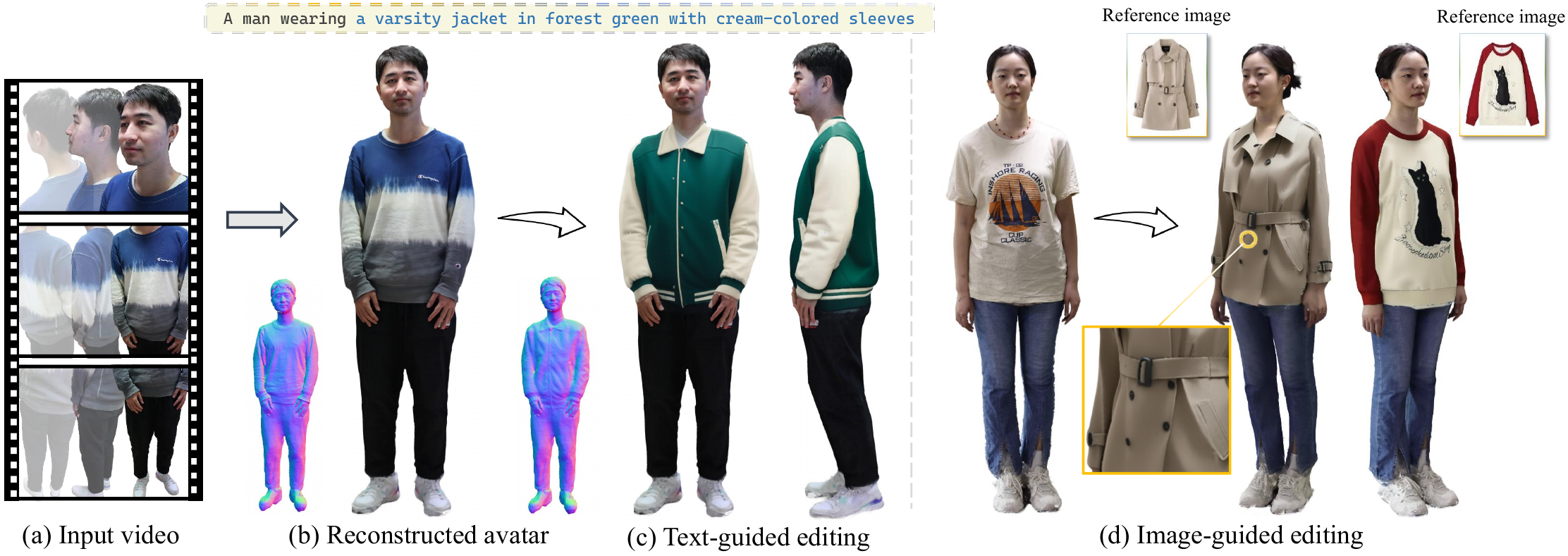}
    \vspace{-0.2cm}
    \captionsetup{type=figure}
    \caption{Given a short RGB video captured by a monocular camera, the corresponding editable 3D avatar can be efficiently generated by our method, achieving both text and image-guided 3D editing with locally adapted geometry and photorealistic renderings.}
    \label{fig:teaser}
\vspace{-0.2cm}
\end{strip}

\vspace{-0.9cm}
\input{sec/0_abstract}  
\vspace{-0.1cm}
\input{sec/1_intro}
\input{sec/2_related}
\input{sec/3_method}
\input{sec/4_exp}

\vspace{-0.1cm}
\section{Conclusion}
\vspace{-0.1cm}
In this paper, we tackled the challenging task of personalized editable 3D avatar creation from real-world videos under the instruction of text prompts or reference images. We demonstrated that the naive solution of direct 3DGS editing leads to degenerated quality, and proposed a novel hybrid tetrahedron-constrained Gaussian Splatting (TetGS) for controllable editing. The proposed TetGS naturally supports decoupled editing, including a localized spatial adaptation module with explicitly partitioned tetrahedrons under delicately designed supervisions, and an appearance generation module that progressively produces coherent texture on the redistributed Gaussians in a coarse-to-fine manner. Both qualitative and quantitative experiments demonstrated that our approach enables high-fidelity photorealistic 3D avatar editing with diverse identities and accessories.

{
    \small
    \bibliographystyle{ieeenat_fullname}
    \bibliography{main}
}

\input{sec/X_suppl}

\end{document}

%% file: sec/0_abstract.tex
\vspace{-0.15cm}
\begin{abstract}
Personalized 3D avatar editing holds significant promise due to its user-friendliness and availability to applications such as AR/VR and virtual try-ons. 
Previous studies have explored the feasibility of 3D editing, but often struggle to generate visually pleasing results, possibly due to the 
unstable representation learning under mixed optimization of geometry and texture in complicated reconstructed scenarios.
In this paper, we aim to provide an accessible solution for ordinary users to create their editable 3D avatars with precise region localization, geometric adaptability, and photorealistic renderings. To tackle this challenge, we introduce a meticulously designed framework that decouples the editing process into local spatial adaptation and realistic appearance learning, utilizing a hybrid Tetrahedron-constrained Gaussian Splatting (TetGS) as the underlying representation. TetGS combines the controllable explicit structure of tetrahedral grids with the high-precision rendering capabilities of 3D Gaussian Splatting and is optimized in a progressive manner comprising three stages: 3D avatar instantiation from real-world monocular videos to provide accurate priors for TetGS initialization; localized spatial adaptation with explicitly partitioned tetrahedrons to guide the redistribution of Gaussian kernels; and geometry-based appearance generation with a coarse-to-fine activation strategy. Both qualitative and quantitative experiments demonstrate the effectiveness and superiority of our approach in generating photorealistic 3D editable avatars.
\end{abstract}

\vspace{-0.1cm}

%% file: sec/1_intro.tex
\vspace{-0.45cm}
\section{Introduction}
\vspace{-0.2cm}
\label{sec:intro}

With the rise of 2D text-to-image generative models~\cite{rombach2022high,saharia2022photorealistic,ramesh2021zero}, the task of text-to-3D avatar creation has garnered significant attention, due to their convenience and the great demand in AR/VR and gaming industries. Many valuable efforts have been made in sculpting iconic 3D characters~\cite{kolotouros2024dreamhuman,huang2024dreamwaltz,huang2024humannorm,liao2024tada} or personalized figures based on high-level user prompts~\cite{zeng2023avatarbooth,huang2024tech,xie2024dreamvton,xiu2024puzzleavatar,men2024en3d}. However, due to their susceptible reliance on diffusion priors~\cite{poole2022dreamfusion,wang2024prolificdreamer}, their generated visual results exhibit a significant gap compared to real-world individuals. Meanwhile, 3D Gaussian Splatting (3DGS)~\cite{kerbl20233d} has demonstrated exceptional capabilities in multi-view reconstruction, achieving photorealistic real-time rendering. However, it is limited to recovering individuals with existing apparel and does not support personalized editing under flexible user control.
What if we could construct and freely edit full-body personalized avatars with high fidelity and photorealism akin to real-world individuals? As shown in Fig.~\ref{fig:teaser}, this paper aims to address the challenging task of creating photorealistic 3D avatars with locally edited accessories, given easily captured real-world monocular videos and user-specified high-level instructions (e.g., texts and reference images). This capability will greatly improve the efficiency of creating controllable 3D digital humans, allowing ordinary users to easily create their unique editable virtual avatars, which can be further applied in industries such as e-commerce for 3D virtual try-on.

To achieve this goal, a straightforward approach is to directly edit the reconstructed 3DGS under random generative guidance~\cite{wang2024gaussianeditor, zhuang2024tip}. However, these methods struggle to produce realistic geometric changes and suffer from blurred renderings with noticeable artifacts, falling far short of the performance in reconstruction tasks. The key technical challenge of 3DGS editing lies in two aspects: On the one hand, the discrete structure of 3DGS restricts gradient propagation to distant points, leading to reliance on the distribution of reconstructed results for editing. This complicates direct 3DGS editing for cases involving significant geometric changes. On the other hand, the densification and optimization processes of 3DGS are sensitive to gradients, particularly in less-constrained editing tasks with diffusion-based generative guidance. In these cases, Gaussians struggle to learn accurate spatial distribution and appearance attributes. Certain methods~\cite{chen2024dge,chen2024gaussianvton,khalid20253dego} utilize multi-view edited images to ensure consistent supervision. However, it is still challenging to maintain 3D consistency in scenes with large viewpoint variations, particularly for $360^{\circ}$ full-body avatars.

To solve the aforementioned challenges, we present \textit{TetGS}, a novel representation where Gaussian kernels are explicitly embedded and updated along with tetrahedron grids. Our key insight lies in the integration of tetrahedral grids and Gaussian Splatting, which allows for more controllable geometric deformation compared to discrete Gaussian kernels.
The structured nature of tetrahedral grids and their direct transition with implicit signed distance field facilitate this controllability. 
Furthermore, the hybrid architecture of TetGS inherently encourages disentangled learning of spatial allocation and texture generation. 
Thereby, we propose to decouple the editing process of 3D Gaussians into localized geometry adaptation and appearance learning, in cooperation with the delicately designed tetrahedron-constrained Gaussian splatting. By leveraging the updates of the structured tetrahedrons to guide the movement and redistribution of Gaussian kernels, we effectively alleviate the unstable spatial optimization caused by the original 3DGS's discrete point cloud-like structure. 
After that, by using few-shot inpainted images for view-consistent supervision in a coarse-to-fine manner, the training of 3D Gaussian focuses on coherent and photorealistic appearance generation.


To this end, we achieve high-precision reconstruction and editing of 3D avatars by utilizing TetGS as a novel hybrid underlying representation. Our approach consists of three stages:
\textit{prior reconstruction} to recover accurate surface and high-fidelity appearance,
\textit{localized spatial adaptation} for flexible and controllable geometric updating, and 
\textit{texture generation} leveraging the superior rendering capabilities of Gaussian splatting under progressive refinement.
Our method supports multiple editing scenarios, including both text-guided editing and reference image-based 3D virtual try-on.
In conclusion, our contributions are threefold:
\begin{itemize}
\item We present a novel system for creating photorealistic 3D editable avatars from easily captured monocular videos by following user-specific prompts. Our proposed pipeline can be seamlessly integrated into applications such as 3D virtual fitting.
\item We introduce a novel hybrid representation named TetGS, which combines the controllable spatial structure of tetrahedral grids with the high-precision rendering capabilities of 3D Gaussian splatting.
\item We decouple the editing phase into localized spatial adaptation and appearance generation. Collaborating with the elaborately designed TetGS, our method achieves coherent $360^{\circ}$ full-body editing with accurate geometry and photorealistic rendering, across diverse editing scenarios.
\end{itemize}



%% file: sec/2_related.tex
\vspace{-0.1cm}
\section{Related Works}
\vspace{-0.1cm}

\begin{figure*}
\vspace{-0.1cm}
\begin{center}
\setlength{\abovecaptionskip}{0.1cm}
\setlength{\belowcaptionskip}{-0.1cm}
\includegraphics[width=1.0\linewidth]{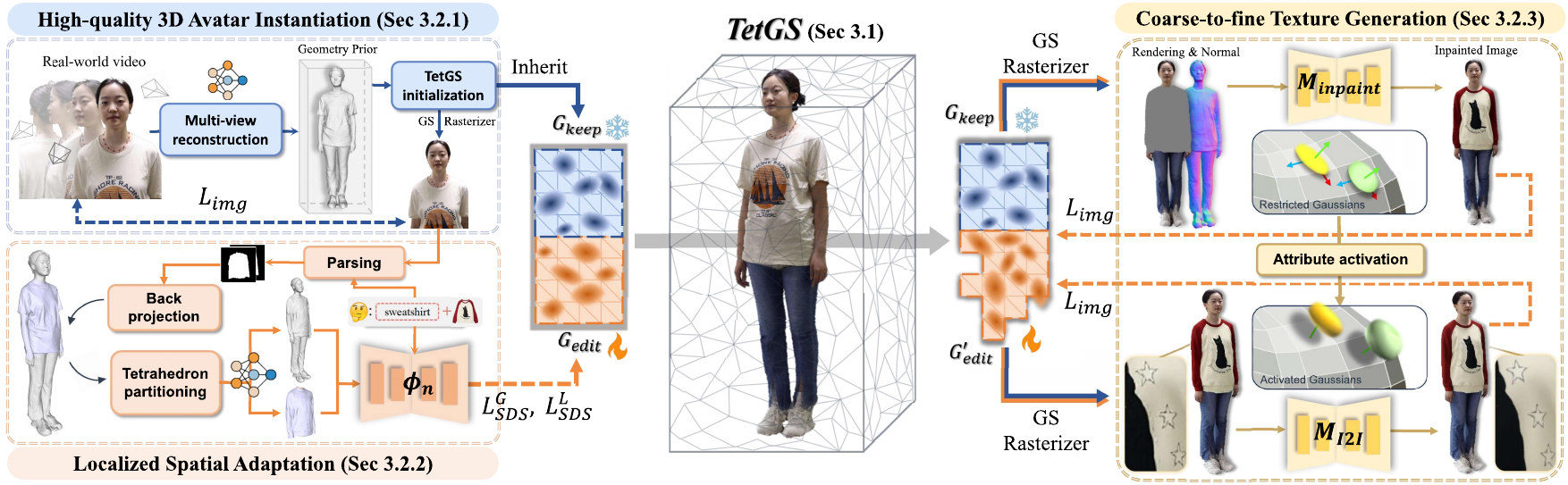}
\caption{An overview of our method, built upon the proposed hybrid Tetrahedron-constrained Gaussian Splatting (TetGS). Our method first learns accurate TetGS initialization from a monocular video, then updates the spatial allocation of localized editing Gaussians along with explicitly partitioned tetrahedrons under diffusion guidance. With the learned distribution, we perform texture editing by optimizing restricted Gaussians with few-shot inpainted images and activating their attributes under augmented guidance.}

\label{fig:network}
\end{center}
\vspace{-0.7cm}
\end{figure*}

{ \noindent \bf 3D Neural Representation. }
3D neural representations have demonstrated great effectiveness in novel-view synthesis, multi-view reconstruction, and 3D surface reconstruction. Neural Radiance Field (NeRF)~\cite{mildenhall2021nerf} and its follow-ups~\cite{muller2022instant, barron2021mip, barron2023zip, barron2022mip, verbin2022ref} achieve high-quality novel-view rendering utilizing neural implicit fields with volume rendering. For accurate surface reconstruction, NeuS~\cite{wang2021neus} and its variants~\cite{fu2022geo, cai2023neuda, li2023neuralangelo, wang2023neus2, wang2023adaptive} use signed distance functions (SDF) that represent 3D surfaces as the zero-level set of SDF. Recently, 3D Gaussian Splatting (3DGS) methods~\cite{kerbl20233d, lyu20243dgsr, guedon2024sugar, cheng2024gaussianpro, huang20242d} have emerged as an efficient representation, offering fast convergence and photorealistic quality. However, these methods fall short in personalized editing, as they can only recover existing captured scenes.

{ \noindent \bf Text to 3D Human Creation. } 
With the rise of 2D diffusion models~\cite{rombach2022high, ruiz2023dreambooth,zhang2023adding,saharia2022photorealistic} and the employment of the SDS loss~\cite{poole2022dreamfusion} in 3D content creation, there have been a large number of works focusing on text-to-3D avatar generation. Some methods~\cite{cao2024dreamavatar,huang2024dreamwaltz,kolotouros2024dreamhuman,xu2023seeavatar,huang2024humannorm,sun2024barbie,liao2024tada} utilize NeRF, DMTet~\cite{shen2021deep}, or parametric SMPL models~\cite{loper2023smpl} for iconic 3D avatar sculpting with input texts. Other methods~\cite{zeng2023avatarbooth,huang2024tech,xiu2024puzzleavatar,men2024en3d} achieve personalized human creation under the condition of both texts and image collections, based on concept-driven 2D diffusion models~\cite{ruiz2023dreambooth,hu2021lora}. Recent works~\cite{liu2024humangaussian,yuan2024gavatar,zhou2024headstudio} also explore text-driven 3D human generation with Gaussian Splatting. Although remarkable efforts have been made, the generated visual quality of these methods still falls short in realism, yielding distorted geometry, over-saturated color, and less detailed appearance.

{ \noindent \bf 3D Scene Editing. }
To combine the advantages of the above two areas, various methods have been proposed to enable 3D reconstructed scene editing with flexible, customized prompts. Instruct-NeRF2NeRF~\cite{haque2023instruct} pioneers in text-guided 3D editing by using Instruct-pix2pix~\cite{brooks2023instructpix2pix} to iteratively update the training datasets, and is followed by a sequence of NeRF-based editing methods~\cite{he2024customize,liu2024genn2n,chen2024consistdreamer,zhuang2023dreameditor,koo2024posterior}. However, the NeRF representation is often limited to global editing due to its implicit parameterization. To enable localized editing and efficient rendering, other methods adopt 3D Gaussian Splatting for its superior rendering fidelity and explicit point data structure. GaussianEditor~\cite{wang2024gaussianeditor} introduces a hierarchical architecture with a semantic tracing strategy to specify the editing region. TIP-Editor~\cite{zhuang2024tip} proposes text-and-image-guided 3D editing for direct appearance control. However, these methods often experience minimal geometric changes, degraded renderings, and affected contents in irrelevant regions, due to the uncontrollable optimization of both geometry and texture under fluctuant generative gradients. For more stable 3DGS editing, certain methods~\cite{chen2024dge,khalid20253dego,chen2024gaussianvton,wang2025view, cao2024gs} leverage consistent multi-view 2D image editing models to guide the adaptation of Gaussian kernels in 3D space. But their consistency decreases in complex editing scenes with a wide range of perspectives, especially when handling $360^{\circ}$ full-body avatars.

%% file: sec/3_method.tex
\section{Method Description}
\vspace{-0.1cm}

Given a $360^{\circ}$ real-world monocular video of a person, we aim to generate a locally editable 3D avatar with high-fidelity and realistic appearance, following user-provided text prompts or reference images.
Fig.~\ref{fig:network} illustrates an overview of our proposed pipeline. Our method relies on a hybrid tetrahedron-constrained Gaussian splatting (TetGS), which we describe in Sec.~\ref{sec:tetgs}. In Sec.~\ref{sec:edit}, we elaborate on our 3D avatar editing pipeline, which utilizes the delicately designed TetGS as the underlying representation to obtain localized high-fidelity editing results.

\subsection{Tetrahedron-constrained Gaussian Splatting}
\label{sec:tetgs}
In this section, we introduce TetGS, a hybrid tetrahedron-constrained 3D Gaussian Splatting where Gaussian kernels are explicitly embedded inside tetrahedral grids. 
We represent the tetrahedron grids $\{V_T, T\}$ following DMTet~\cite{shen2021deep}, where an MLP $\psi_g(x)$ predicts the signed distance function (SDF) of grid points $V_T$ on tetrahedrons $T$. To insert Gaussians inside this structured grid,
we dive into the underlying structure of tetrahedral meshes. When applying the Marching Tetrahedron (MT) algorithm~\cite{doi1991efficient} that enables explicit triangular mesh extraction $M= \{\mathcal{V}, \mathcal{F}\}$ from $\{V_T, T\}$, each mesh vertex $v_{ab}^M \in \mathcal{V}$ is extracted from a unique tetrahedral edge $e_{ab}^T = \{v_{a}^T, v_{b}^T\}$ with the sign change, by linearly interpolating the SDF values of its two endpoints: 
\begin{equation}
\ v_{ab}^M = \frac{v_a^T \cdot s(v_b^T)-v_b^T \cdot s(v_a^T)}{s(v_b^T)-s(v_a^T)}.
\end{equation}
Thus, we can define the mapping from mesh vertex to valid tetrahedral edge as $g_{v \rightarrow e}: v_{ab}^M \rightarrow e_{ab}^T = \{v_a^T, v_b^T\}$. 
Since each mesh triangle is extracted inside a single tetrahedron in the surface configuration of MT, for a triangle $f_i = \{v_{i_1}^M, v_{i_2}^M, v_{i_3}^M\} \in \mathcal{F}$, we can obtain the mapping $h_{f \rightarrow t}: f_i \rightarrow t_k$ to its father tetrahedron $t_k \in T$, where $t_k$ satisfies $\{g_{v \rightarrow e}(v_{i_1}^M), g_{v \rightarrow e}(v_{i_2}^M), g_{v \rightarrow e}(v_{i_3}^M)\} \subseteq t_k$.
By allocating Gaussians onto each mesh triangle $f_i$, we naturally embed Gaussian kernels inside the father tetrahedron $t_k$. The position $\mu$ of each Gaussian is calculated from the vertex coordinates $v^M$ of its occupied mesh triangle $f_i$ with predefined barycentric weights $w$ : $\mu = w_av_{i_1}^M + w_bv_{i_2}^M + w_cv_{i_3}^M + \tau \mathbf{n}$, where $\tau$ is a learnable displacement along the surface normal $\mathbf{n}$ and is set to zero at the beginning. We define the remaining Gaussian attributes following the original 3DGS, including a 3D covariance matrix $\sum \in \mathbb{R}^7$, SH coordinates $c \in \mathbb{R}^k$, and an opacity $\sigma \in \mathbb{R}$.

\begin{figure}
\begin{center}
\setlength{\abovecaptionskip}{0.1cm}
\setlength{\belowcaptionskip}{-0.1cm}
\includegraphics[width=0.75\linewidth]{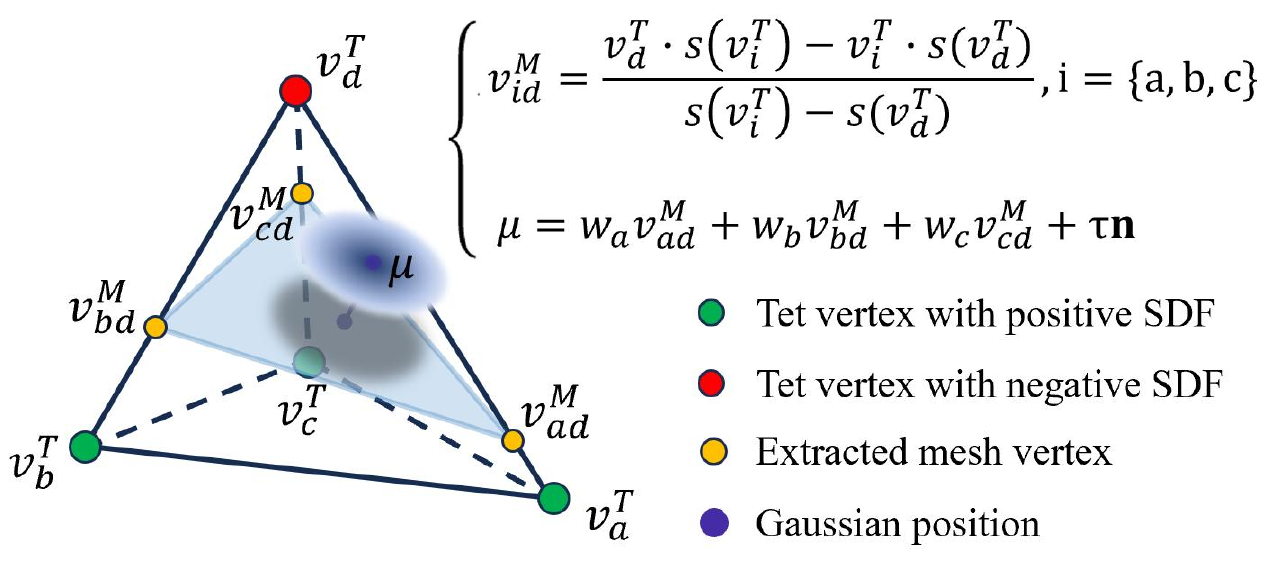}
\caption{An illustration of tetrahedron-constrained Gaussian. Each Gaussian kernel is embedded in a unique tetrahedron with its position $\mu$ calculated from the SDF of tetrahedral vertices.}

\label{fig:tetgs}
\end{center}
\vspace{-0.7cm}
\end{figure}

At this point, we have established a direct correlation between the tetrahedron $t_k$ and its embedded Gaussians $\mathcal{G}_k$, where $\mathcal{G}_k$ are all the Gaussians assigned on mesh triangles extracted from $t_k$ (empty set for invalid tetrahedrons) and the position of $\mathcal{G}_k$ is directly determined by the SDF values of the tetrahedral vertices, as illustrated in Fig.~\ref{fig:tetgs}. We record this relation using a new Gaussian attribute $k$ indicating the index of its father tetrahedron.
By updating the SDF of the tetrahedral vertices, we enable structured spatial guidance for the dynamic movement and redistribution of Gaussian kernels during editing. This also enables localized maintenance for non-editing Gaussians through explicit tetrahedron partitioning.

\subsection{3D Avatar Editing with TetGS}
\label{sec:edit}
With the designed TetGS, we propose our 3D avatar editing pipeline which consists of three modules: \textit{high-quality 3D avatar instantiation, localized spatial adaptation, and coarse-to-fine appearance generation.} We first recover accurate geometry and appearance prior from monocular videos for TetGS initialization (Sec~\ref{sec:recon}). Building on this, we introduce a localized spatial adaptation strategy through explicit partitioning of tetrahedral grids (Sec~\ref{sec:geo}). The optimized tetrahedrons provide direct guidance for the reallocation of the editing Gaussians and enable few-shot normal-based texture inpainting on restricted Gaussian disks, followed by attribute activation with augmented guidance to capture high-frequency details (Sec~\ref{sec:tex}). Additionally, with the controllable TetGS, we also achieve reference image-based 3D virtual try-on (Sec~\ref{sec:ref}).

\vspace{-0.0cm}
\subsubsection{High-quality 3D avatar Instantiation}
\label{sec:recon}
This module aims to construct high-quality 3D avatars with precise surface and detailed appearance, given real-world monocular videos.
We represent the 3D avatar in an SDF field instantiated by an MLP $\psi(x):  \mathbb{R}^3\to \mathbb{R}$, following previous surface reconstruction methods~\cite{wang2021neus}.
To construct a smooth surface with detailed human features, we adopt the deformable anchors with positional encoding proposed in NeuDA~\cite{cai2023neuda}. 
To further tackle the irregular holes caused by uneven light and extreme reflection in real-world data, we adopt a normal regularization loss~\cite{verbin2022ref} by constraining the gradient $\hat{\mathbf{n}}_i$ of SDF with the predicted normal $\hat{\mathbf{n}}'_i$:
\vspace{-0.1cm}
\begin{equation}
\ L_p = \sum_i w_i ||\hat{\mathbf{n}}_i - \hat{\mathbf{n}}'_i||^2, 
\vspace{-0.2cm}
\end{equation}
where $w_i$ is the weight of the $i$th sample point on a ray. We also apply a normal orientation loss~\cite{verbin2022ref} to alleviate surface depression along the ray direction $\hat{\mathbf{d}}$:
\vspace{-0.1cm}
\begin{equation}
\ L_o = \sum_i w_i \max(0, \hat{\mathbf{n}}_i \cdot \hat{\mathbf{d}})^2. 
\vspace{-0.2cm}
\end{equation}

After surface reconstruction, we can initialize the TetGS by converting the reconstructed geometry into tetrahedron grids and optimizing their embedded Gaussians $\mathcal{G}$ with the multi-view pixel loss. This high-fidelity TetGS initialization provides accurate priors for the later 3D avatar editing.

\subsubsection{Localized spatial adaptation}
\label{sec:geo}
At the core of TetGS-based editing, we perform localized spatial adaptation in editable tetrahedrons to guide the redistribution of Gaussian kernels.
Ideally, a localized editing task should meet two conditions: 1) the editing region can be freely changed without being restricted to the original shape, and 2) the parameters of irrelevant Gaussians stay identical to the reconstructed ones. We achieve this by explicitly partitioning the tetrahedrons into two subsets and derive a localized spatial editing scheme.

{ \noindent \bf Localized tetrahedron partitioning. }
We start from explicit mesh localization and gradually deduce 
into the tetrahedron space.
Due to the structured nature of the extracted MT mesh $M$, we can directly divide the mesh triangles by back-projecting multi-view masks of the interested area.
Masks can be produced by existing segmentation~\cite{kirillov2023segment} or human parsing models~\cite{li2020self}. Mesh triangles corresponding to preserved and editing regions are denoted as $\mathcal{F}^{keep}$ and $\mathcal{F}^{edit}$, respectively. Unlike the previous work~\cite{zhuang2024tip} that manually defines the editing scope, we instead localize the areas of the preserved Gaussians and treat all the remaining regions as editable to allow maximal editing effect. We find the father tetrahedrons $T^{keep}$ of triangles $\mathcal{F}^{keep}$ through the mapping $h_{f \rightarrow t}$. The preserved Gaussians $\mathcal{G}^{keep}$ are defined as Gaussians with their father indices $k$ pointing to $T^{keep}$.

\begin{figure}
\begin{center}
\setlength{\abovecaptionskip}{0.1cm}
\setlength{\belowcaptionskip}{-0.1cm}
\includegraphics[width=0.99\linewidth]{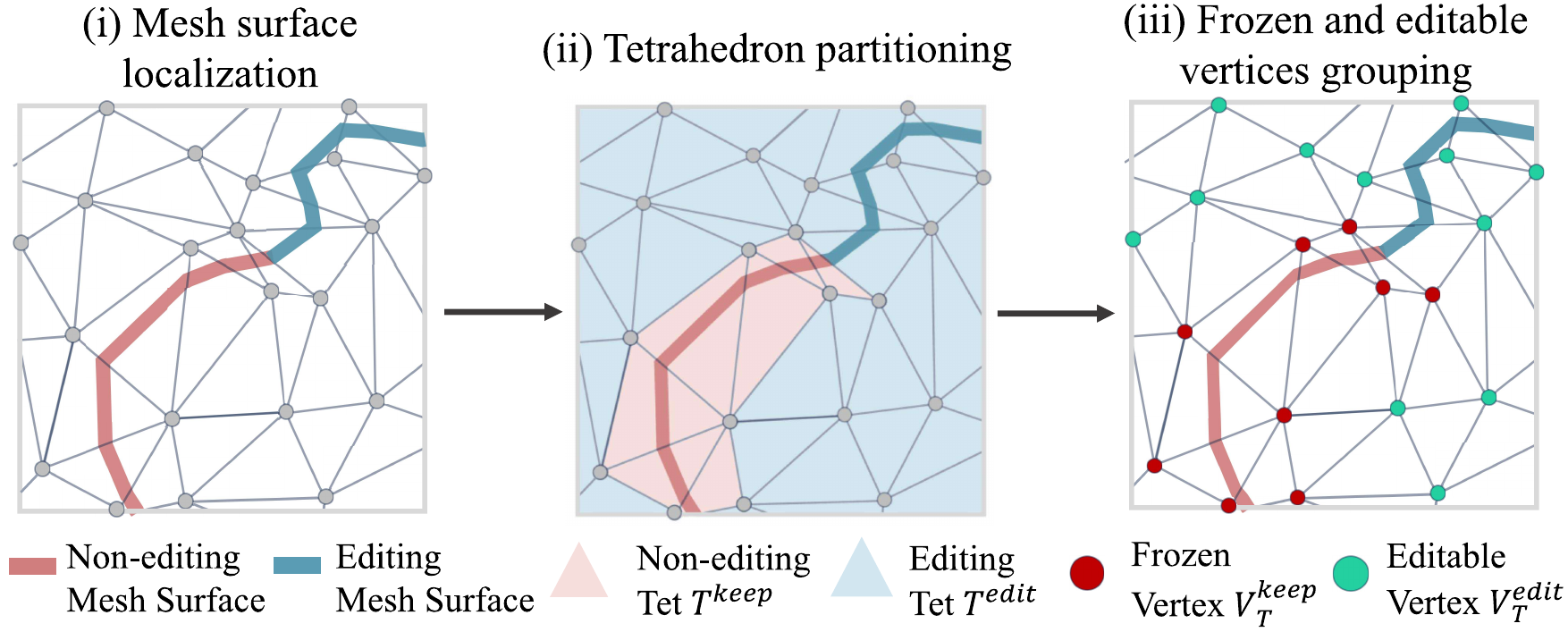}
\caption{Demonstration of the tetrahedron partitioning process. (i) Localized mesh triangles given multi-view mask labels. (ii) Tetrahedron partitioning according to their extracted triangles. (iii) Tetrahedral vertices grouping into frozen and editable ones.}

\label{fig:tet}
\end{center}
\vspace{-0.5cm}
\end{figure}

To ensure the strict spatial maintenance of $\mathcal{G}^{keep}$, we constrain the spatial information of the father tetrahedrons $T^{keep}$ by freezing the SDF values of their vertices $V_T^{keep}$ during the spatial adaptation stage. We update the SDF of the remaining vertices $V_T^{edit} = \{v^T \notin V_T^{keep} \mid v^T \in V_T\}$ in editing
tetrahedrons $T^{edit} = \{t \notin T^{keep} \mid t \in T\}$ with the MLP $\psi_g$ to encourage flexible geometric variation. A demonstration of this process can be found in Fig.~\ref{fig:tet}.

{ \noindent \bf Dual spatial constraint. }
Based on the explicit structure of tetrahedron grids, the spatial adaptation of TetGS can be directly supervised with the SDS loss~\cite{poole2022dreamfusion} on the normal renderings based on a pre-trained T2I diffusion model $\phi_n$ under the text condition $y^G$. Given the global surface normal $\mathbf{n}_G = N(M', p)$, where $p$ is the camera parameter and $M'$ denotes the intermediate MT mesh, we define $L_{SDS}^G$ as:
\vspace{-0.1cm}
\begin{equation}
\ L_{SDS}^G = \mathbb{E}_{t,\epsilon} \left[ w_t(\hat{\epsilon}_{\phi_n}(\mathbf{z}_t^{\mathbf{n}_G};y^G,t)-\epsilon)\frac{ \partial \mathbf{n}_G }{ \partial \psi_g }\frac{ \partial \mathbf{z}^{\mathbf{n}_G} }{ \partial \mathbf{n}_G } \right].
\vspace{-0.1cm}
\end{equation}
However, in the localized editing task, the supervision focus of the global and local regions significantly differs, as the global region ensures overall harmony regarding the style of the preserved content, while the local editing region is encouraged to produce distinct and detailed changes.
Thus, with the explicitly partitioned editing tetrahedrons $T^{edit}$, we further design a local SDS loss focusing on the spatial updating of the editing geometry $M'_{edit} = MT(T^{edit})$:
\vspace{-0.1cm}
\begin{equation}
\ L_{SDS}^L = \mathbb{E}_{t,\epsilon} \left[w_t(\hat{\epsilon}_{\phi_n}(\mathbf{z}_t^{\mathbf{n}_L};y^L,t)-\epsilon)\frac{ \partial \mathbf{n}_L }{ \partial \psi_g }\frac{ \partial \mathbf{z}^{\mathbf{n}_L} }{ \partial \mathbf{n}_L } \right].
\vspace{-0.1cm}
\end{equation}
The dual constraint of global and local regions helps to ensure harmonious overall changes and enables precise control over the editing content.

{ \noindent \bf Surface regularizations. }
The above tetrahedron partitioning strategy enables the unchanged spatial distribution of the non-editing Gaussians $\mathcal{G}^{keep}$ by freezing the SDF of $V_T^{keep}$. However, since the SDF of the editing vertices $V_T^{edit}$ is updated with the continuous MLP $\psi_g$, the naive solution of optimizing $\psi_g$ without constraint can lead to the predicted editing surface gradually occluding the preserved $\mathcal{G}^{keep}$. To alleviate this issue, we introduce a surface-aware regularization on the predicted SDF at $V_{T}^{keep}$:
\vspace{-0.15cm}
\begin{equation}
\ L_{sa} = \sum_{v_i \in V_T^{keep}}||\psi_g(v_i) - \hat{s}_i||^2.
\vspace{-0.15cm}
\end{equation}
This term enables $\psi_g$ to be aware of the position of $\mathcal{G}^{keep}$ by constraining the predicted SDF at $v_i \in V_{T}^{keep}$ with the frozen SDF $\hat{s}_i$.
We also apply an additional normal consistency loss $L_{nc}$ for a smoother surface. The overall training objective can be formulated as:
\begin{equation}
\ L = \lambda_{SDS}^G L_{SDS}^G + \lambda_{SDS}^L L_{SDS}^L + \lambda_{sa} L_{sa} + \lambda_{nc} L_{nc}.
\end{equation}

The localized adapted tetrahedrons effectively guide the spatial redistribution of the Gaussian kernels. For the intentionally preserved tetrahedrons $T^{keep}$, we directly assign their Gaussians as $\mathcal{G}^{keep}$ inherited from the reconstructed avatar. For the editing tetrahedrons $T^{edit}$ with deformed geometry, we initialize and reallocate their Gaussians $\mathcal{G}^{edit}$ by following the procedure described in Sec~\ref{sec:tetgs}.

\subsubsection{Texture generation within editing regions}
\label{sec:tex}
\vspace{-0.0cm}
For the reallocated editing Gaussians $\mathcal{G}^{edit}$ with already learned spatial information, we aim to generate photorealistic texture in a progressive manner, where we learn a coarse appearance with the restricted TetGS using few-shot inpainted images and further activate their attributes with augmented multi-view renderings.

{ \noindent \bf Restricted TetGS under few-shot supervision. }
Inspired by~\cite{richardson2023texture,zeng2024paint3d,wu2023hyperdreamer}, we utilize a normal-based inpainter for few-shot coarse appearance learning of $\mathcal{G}^{edit}$ to achieve stable training and higher rendering quality. To complement the less-supervised texturing, we restrict partial attributes of $\mathcal{G}^{edit}$ onto 2D surfels.
Specifically, we deactivate the position displacement $\tau$ along the normal direction, and degrade the 3D covariance matrix to 2D plane following~\cite{waczynska2024games, lin2024direct}, where Gaussian kernels are simplified as 2D disks covering the entire geometric surface.  
We also fix the opacity $\alpha$ to a constant and simplify the SH coordinates into view-independent $rgb$ to improve rendering robustness in novel views under few-shot guidance.


{ \noindent \bf Normal-guided coarse texture inpainting. }
Based on the restricted TetGS, we learn the coarse editing texture by iteratively sampling different viewpoints and optimizing $\mathcal{G}^{edit}$ under the supervision of the inpainted images. We record the uncolored area with a triangle label on the edited mesh $M'_{edit}$, denoted as $\mathcal{L}: \mathcal{F}’_{edit} \rightarrow \{0, 1\}$, where $\mathcal{F}’_{edit} \rightarrow \{1\}$ initially.
In each iteration $k$, for the sampled view $p_{k}$ focusing at the editing region, we obtain its current rendered image $\hat{I}_{k}$, normal map $\mathbf{n}_{k}$, and uncolored mask $m_{k}$ by $R: (\mathcal{G}^{keep}, \mathcal{G}_{k-1}^{edit}, \mathcal{L}_{k-1}, p_{k}) \rightarrow (\hat{I}_{k}, \mathbf{n}_{k}, m_{k})$. 
A pre-trained normal-based inpainter $\mathcal{M}$ is used to fill the uncolored area and generate coherent inpainted image $I_{k}^{paint} = \mathcal{M}(\hat{I}_{k}, \mathbf{n}_{k}, m_{k})$, with clues provided by the existing appearance. We blend $I_{k}^{paint}$ and the original rendering $\hat{I}_{k}$ with a blurred blending mask $m_{k}^b = Blur(m_{k})$, and obtain the guidance image $I_k$, which is formulated as:
\vspace{-0.1cm}
\begin{equation}
\ I_{k} = m_{k}^b \odot I_{k}^{paint} + (1-m_{k}^b) \odot \hat{I}_{k}.
\vspace{-0.1cm}
\end{equation}

Using $I_{k}$ as the supervision of the current view, we optimize trainable editing Gaussians $\mathcal{\hat{G}}_{k}^{edit}$ under the view $p_{k}$ via the pixel-level reconstruction loss. $\mathcal{\hat{G}}_{k}^{edit}$ is a subset of $\mathcal{G}^{edit}_k$ and is defined as Gaussians assigned on the masked triangles obtained by back-projecting the binarized $m_{k}^b$ onto $M'_{edit}$. The training objective for the current view $p_{k}$ can be formulated as:
\begin{equation}
\ L(\mathcal{\hat{G}}_{k}^{edit}; \hat{I}_{k}, I_k) = L_1(\hat{I}_{k}, I_k) + \lambda L_{SSIM}(\hat{I}_{k}, I_k).
\end{equation}
After the optimization of the $k$th iteration, the Gaussians $\mathcal{G}^{edit}_k$ and triangle label $\mathcal{L}_{k}$ are updated with $\mathcal{\hat{G}}_{k}^{edit}$ and the corresponding masked triangles, respectively.

In practice, due to the uniform shape of the human body, we first inpaint front and back views simultaneously for consistent global style, while local consistency is ensured by the normal-based inpainter, with which we progressively sample cameras around the avatar under different elevations and azimuths covering the whole editing region.

{ \noindent \bf Attribute activation with augmented guidance. }
The above-edited appearance shows coherence across different views and with preserved regions, but still suffers from degenerated rendering quality under novel views due to the few-shot supervision with the restricted TetGS, as depicted in Fig.~\ref{fig:texture} (b). To fully exploit the rendering potential of Gaussian splatting and further boost the realism of the editing result, we propose to refine $\mathcal{G}^{edit}$ with attribute activation under enhanced guidance. 

To better capture high-frequency details, we release the restrictions on $\mathcal{G}^{edit}$ by relaxing position and covariance to 3D space, allowing for trainable opacity, and converting color back to view-dependent SH coordinates. To remove artifacts and enhance local details, we follow~\cite{meng2021sdedit} and apply an I2I pipeline to obtain refined results $I'_i$ on randomly sampled renderings. Then the editing task can be converted into a reconstruction task with multi-view images $\{I'_i\}_{i=1}^{n}$, where $\mathcal{G}^{edit}$ is optimized with the multi-view image loss.

\subsubsection{Editing with reference images}
\label{sec:ref}
Benefiting from the controllable TetGS, we also achieve 3D virtual try-on given reference garment images. We adopt IDM-VTON~\cite{choi2024improving} to generate front and back-view try-on images $I_f$ and $I_b$ in place of images inpainted by $\mathcal{M}$. We also apply geometric supervision of $I_f$ and $I_b$ during the localized spatial adaptation:
\begin{equation}
\ L_{vton} = \lambda_{norm} ||\mathbf{n}_{\{f, b\}} - \hat{\mathbf{n}}_{\{f, b\}}||^2 + \lambda_{mask} ||m_f - \hat{m}_f||^2 ,
\end{equation}
where $\mathbf{n}_{\{f, b\}}$ and $\hat{\mathbf{n}}_{\{f, b\}}$ denote the rendered and estimated~\cite{saito2019pifu} normals of the front and back views, $m_f$ and $\hat{m}_f$ denote the rendered and foreground masks of the front view, respectively. Thus, both the garment style and geometric design can be transferred into the edited 3D avatar.

%% file: sec/4_exp.tex
\begin{figure*}
\vspace{-0.1cm}
\begin{center}
\setlength{\abovecaptionskip}{0.cm}
\setlength{\belowcaptionskip}{-0.2cm}
\includegraphics[width=0.93\linewidth]{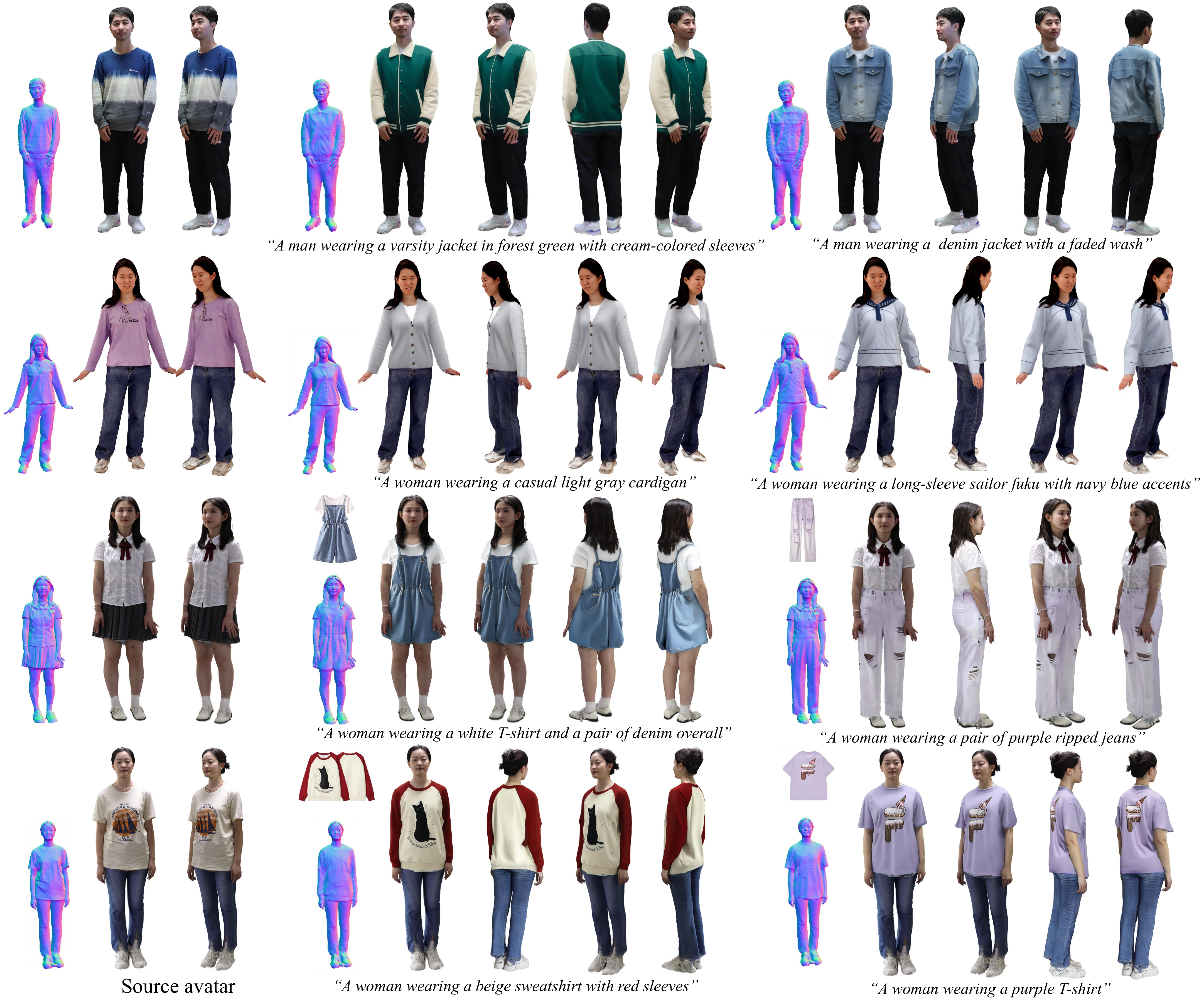}
\caption{Multi-view renderings and the underlying geometries before and after editing with various subjects and accessories.}

\label{fig:res}
\end{center}
\vspace{-0.6cm}
\end{figure*}

\section{Experimental Results}

{ \noindent \bf Implementation details. }
We define TetGS within $512^3$ tetrahedron grids embedded around a standard human body following~\cite{huang2024humannorm,huang2024tech}. A different number of Gaussians are assigned to each tetrahedron based on the face area of its extracted triangles. For 3D avatar instantiation, we use the Adam optimizer~\cite{kingma2014adam} with a learning rate of 1e-3 to train the SDF field $\psi(x)$ and optimize Gaussians $\mathcal{G}$ similar to the original 3DGS method~\cite{kerbl20233d}. For localized spatial adaptation, we update the SDF of tetrahedral vertices using the AdamW optimizer~\cite{loshchilov2017decoupled} with a learning rate of 2e-5 for 10000 iterations, taking around 1.2 hours on a single NVIDIA A40. The loss weights $\{\lambda_{SDS}^G, \lambda_{SDS}^L, \lambda_{sa}, \lambda_{nc}\}$ are set to $\{0.5, 0.5, 5000, 2000\}$. For texture generation, we apply SDXL-based ControlNetPlus~\cite{controlnetplus} for both normal-based inpainting and I2I augmentation. We sample views by rotating the cameras around the editing region with an interval of $30^{\circ}$. Texture inpainting and attribute activation stages take 20 mins and 3 mins for training, respectively.

{ \noindent \bf Datasets. }
To evaluate our method, we create a $360^{\circ}$ captured full-body human dataset containing 10 static monocular videos with different subject identities and outfits. The videos are captured with portable cameras under daily environmental settings. The subjects remain still when we rotate the camera around the heads, upper bodies, and lower bodies for more fine-grained details. All videos have a duration of 40 to 50 seconds with a resolution of $1080\times 1920$, and are split to 250 frames as training data. We use COLMAP~\cite{schoenberger2016sfm} for camera pose estimation. For each editing task, we generate text prompts with GPT-4~\cite{achiam2023gpt} and collect reference images from e-commerce websites.

\subsection{3D editable avatar generation}
Fig.~\ref{fig:res} showcases several editing results, including text-guided editing in rows 1-2 and reference image-based editing in rows 3-4. We illustrate multi-view renderings and the underlying geometries before and after editing with various accessories, achieving reasonable geometric change and photorealistic generated texture coherent with the preserved appearance. More results are shown in the supplementary.

\subsection{Comparison}
{ \noindent \bf Baselines. }
We compare our method with three baselines: GaussianEditor~\cite{wang2024gaussianeditor} and DGE~\cite{chen2024dge}, which are state-of-the-art methods for 3DGS scene editing with text prompts, and TIP-Editor~\cite{zhuang2024tip}, which focuses on image-guided 3DGS editing for specific appearance control. We train all models using the officially released code with default configurations on our collected dataset. For GaussianEditor and DGE, since they use Instruct-pix2pix~\cite{brooks2023instructpix2pix} to update the intermediate rendered images, we convert our text prompts to meet the required format. For TIP-Editor, we use the same reference images as ours and replace the prompts of the editing subjects with special tokens.

{ \noindent \bf Qualitative comparison. }
We demonstrate qualitative comparisons in Fig.~\ref{fig:comp}. GaussianEditor and DGE generate needle-like artifacts and blurred renderings due to the unstable 3DGS optimization and densification under generative guidance. Additionally, they exhibit noticeable color leakage into irrelevant regions because of the accumulated errors during the semantic segmentation process and inconsistent edited guidance under different perspectives. TIP-Editor fails to preserve detailed texture faithful to the reference garment. Moreover, it suffers from noisy appearances and minor geometric changes owing to the random gradients produced by the diffusion priors. In contrast, our method generates high-fidelity edited avatars with reasonable geometric adaptation and coherent renderings, benefiting from the proposed TetGS. We also achieve accurate textures for image-based editing by directly optimizing Gaussians under the guidance of 2D try-on images.

\vspace{-0.1cm}
{ \noindent \bf Quantitative comparison. }
For quantitative evaluation, we use Frechet Inception Distance (FID)~\cite{heusel2017gans} to access the quality of edited images, where a lower FID indicates greater similarity to the reconstructed images, reflecting higher rendering fidelity and realism. To evaluate the alignment of edited avatars with input text prompts, we employ CLIP Text-Image similarity~\cite{radford2021learning}. For the image-guided method TIP-Editor, we also calculate DINO similarity~\cite{oquab2023dinov2} between the reference image and the edited results. All metrics are computed on 60 rendered images captured evenly around the 3D avatar. As shown in Tab.~\ref{tab:IP}, our method outperforms other baselines across all metrics, notably achieving a significant improvement in FID and producing photorealistic results comparable to real-world individuals.

\begin{table}
\setlength{\abovecaptionskip}{0.1cm}
  \centering
    \caption{Quantitative comparison with 3D avatar editing methods on FID, CLIP and DINO. $\uparrow$, $\downarrow$ denote if higher or lower is better.}
      \resizebox{0.8\linewidth}{!}{
  \begin{tabular}{cccccccc}
    \toprule
    Method  & GaussianEditor & DGE & TIP-Editor & Ours \\
    \midrule
    FID $\downarrow$  & 194.54 & 201.82 & 258.27 & \bf{115.95}  \\
    CLIP $\uparrow$  & 22.14 & 23.05  & 22.51 & \bf{26.28} \\
    DINO $\uparrow$  & -   & -  & 0.726 & \bf{0.752} \\

    \bottomrule
  \end{tabular}}
  \label{tab:IP}
  \vspace{-0.4cm}
\end{table}

\begin{figure*}
\vspace{-0.2cm}
\begin{center}
\setlength{\abovecaptionskip}{0.0cm}
\setlength{\belowcaptionskip}{-0.1cm}
\includegraphics[width=0.96\linewidth]{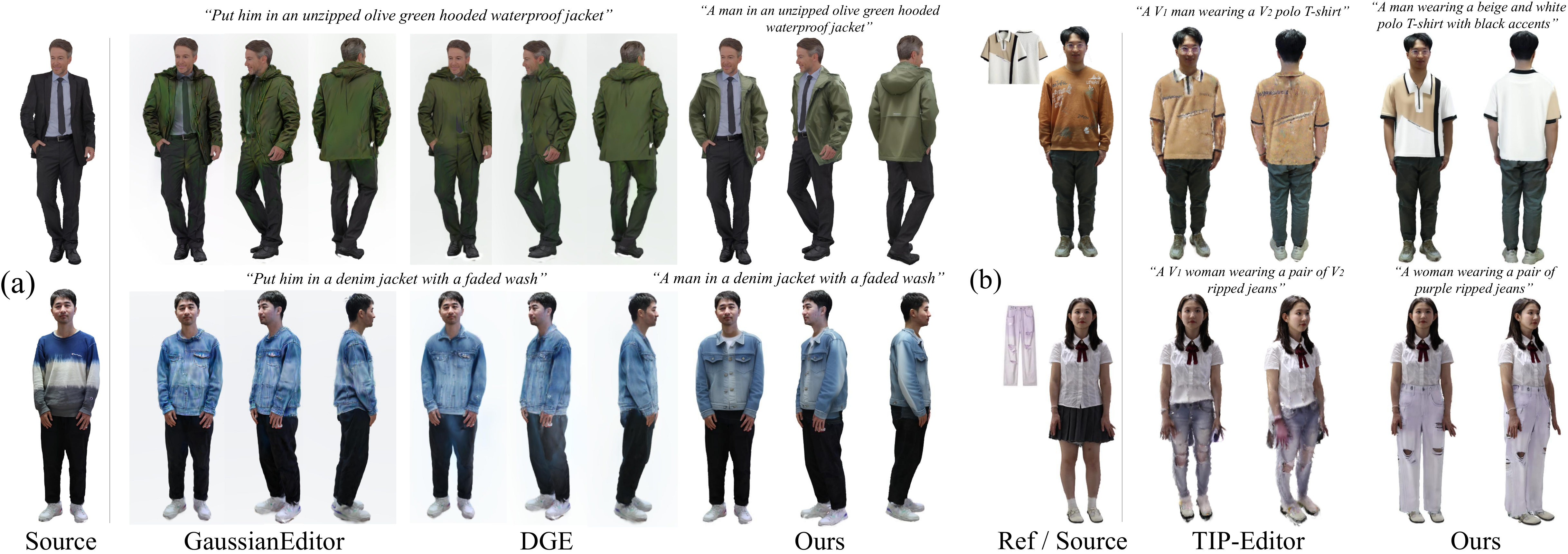}
\caption{Qualitative comparison with text-guided methods GaussianEditor~\cite{wang2024gaussianeditor} and DGE~\cite{chen2024dge}, and image-guided method TIP-Editor~\cite{zhuang2024tip}.}

\label{fig:comp}
\end{center}
\vspace{-0.7cm}
\end{figure*}

\vspace{-0.1cm}
\subsection{Ablation study}
\label{sec:ablation}
\begin{figure}
\begin{center}
\setlength{\abovecaptionskip}{-0.0cm}
\setlength{\belowcaptionskip}{-0.3cm}
\includegraphics[width=0.97\linewidth]{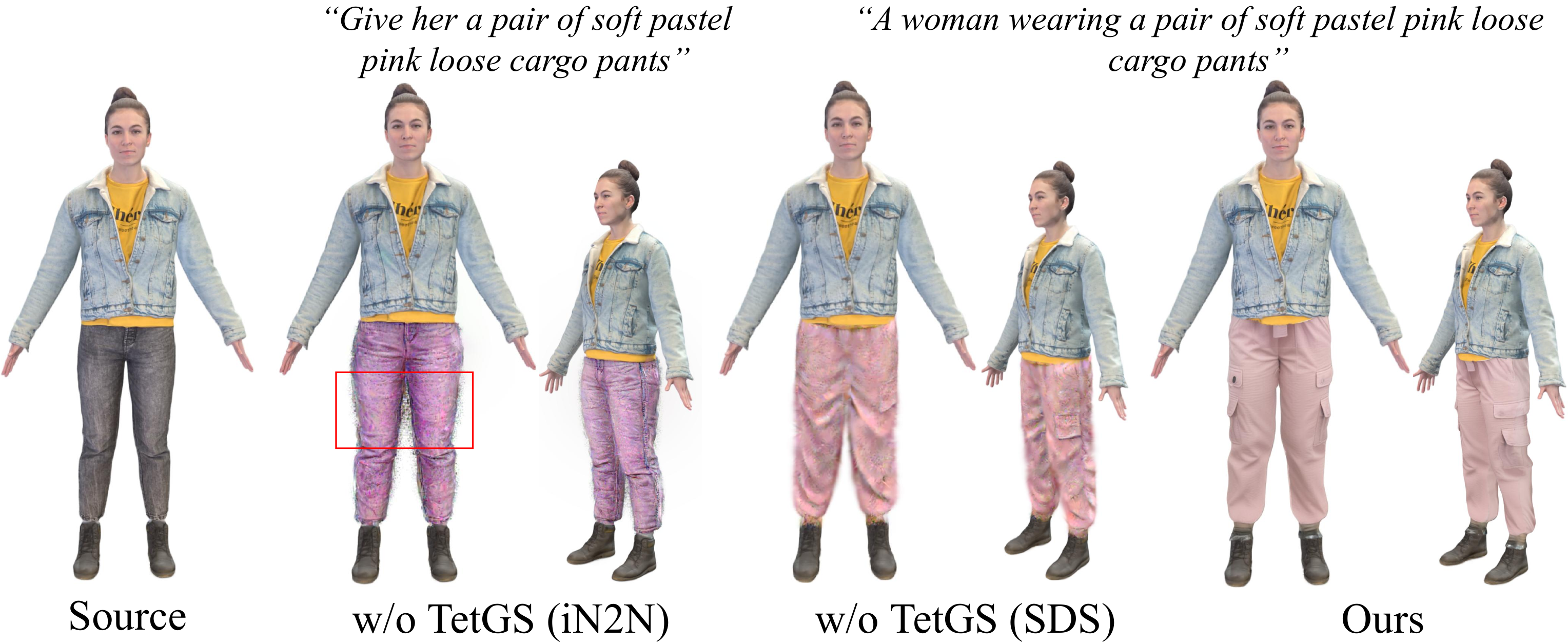}
\caption{Ablation study on decoupled editing with TetGS.}

\label{fig:tetgs_ablation}
\end{center}
\vspace{-0.3cm}
\end{figure}

{ \noindent \bf Ablation on decoupled editing with TetGS. }
Fig.~\ref{fig:tetgs_ablation} and Tab.~\ref{tab:ablation} validate the effect of the TetGS-based decoupled editing. We train two variants that directly optimize all the attributes of $\mathcal{G}_{edit}$ without structured tetrahedrons under the generative supervision of iN2N~\cite{haque2023instruct} and the SDS loss~\cite{poole2022dreamfusion}, respectively. Due to the stochastic diffusion process, the former generates chaotic noises scattering around the editing region, while the latter produces blurry and diverging shape boundaries. They both produce noisy appearances with few details. In contrast, our method achieves geometric changes with clear boundaries and rich details, as well as high-quality renderings.

\begin{table}
\setlength{\abovecaptionskip}{0.1cm}
\setlength{\belowcaptionskip}{0.0cm}
  \centering
    \caption{Ablation study on the proposed editing pipeline. Results verify the effectiveness of the proposed modules.}
      \resizebox{1.0\linewidth}{!}{
  \begin{tabular}{cccccccc}
    \toprule
    Variants  & w/o TetGS (iN2N) &w/o TetGS (SDS) & w/o AA & full model \\
    \midrule
    FID $\downarrow$  & {215.56}  & {217.15} & {120.33} & \bf{115.95} \\
    CLIP $\uparrow$  & {22.64}  & {23.96} & {26.03} & \bf{26.28} \\

    \bottomrule
  \end{tabular}}
  \label{tab:ablation}
\vspace{-0.5cm}
\end{table}

\begin{figure}
\begin{center}
\setlength{\abovecaptionskip}{0.0cm}
\setlength{\belowcaptionskip}{-0.1cm}
\includegraphics[width=0.88\linewidth]{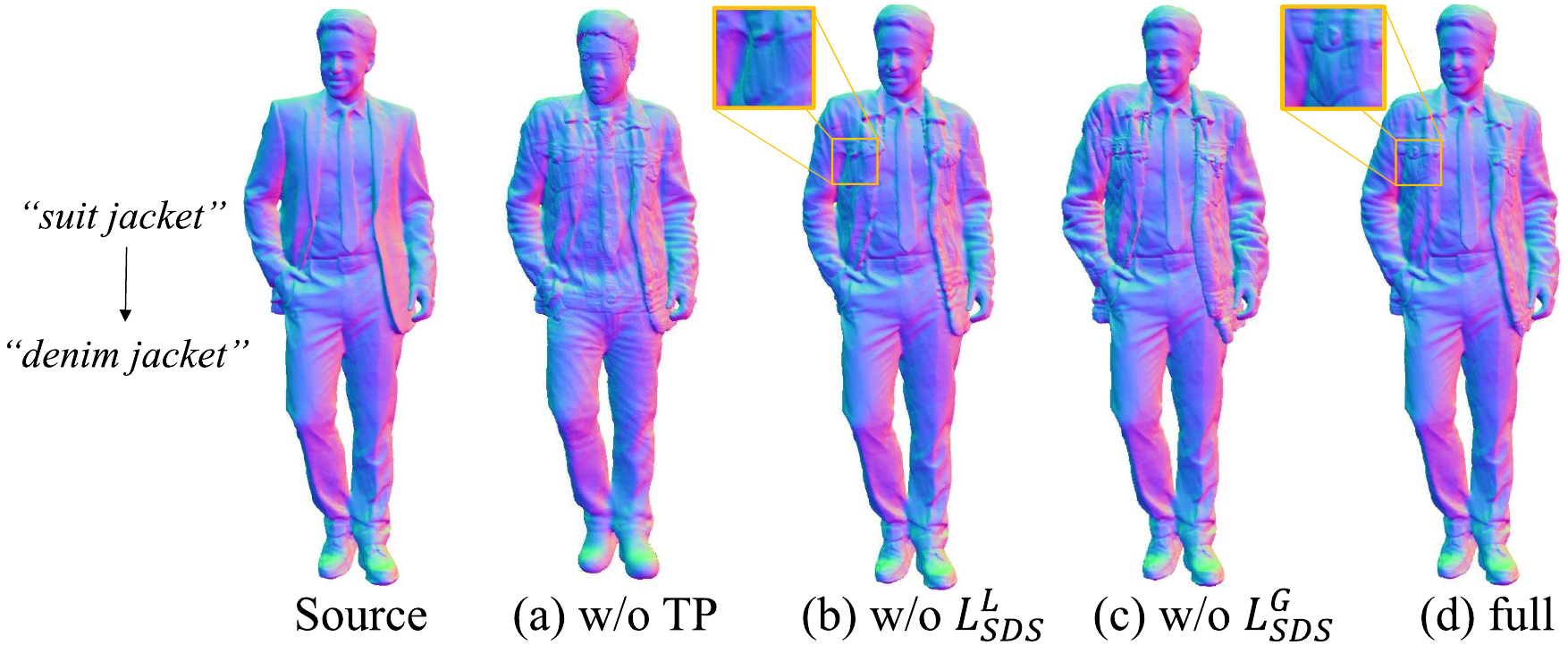}
\caption{Ablation study on localized spatial adaptation.}

\label{fig:spatial}
\end{center}
\vspace{-0.5cm}
\end{figure}

{ \noindent \bf Localized spatial adaptation with dual constraint. }
We verify the designed schemes during the localized spatial adaptation in Fig.~\ref{fig:spatial}. Variants without tetrahedron partitioning (TP), local SDS loss $L_{SDS}^L$, and global SDS loss $L_{SDS}^G$ result in unintended modifications to non-editing regions, blurry details, and unnatural edited shapes, respectively.

\begin{figure}
\begin{center}
\setlength{\abovecaptionskip}{0.cm}
\setlength{\belowcaptionskip}{-0.1cm}
\includegraphics[width=0.81\linewidth]{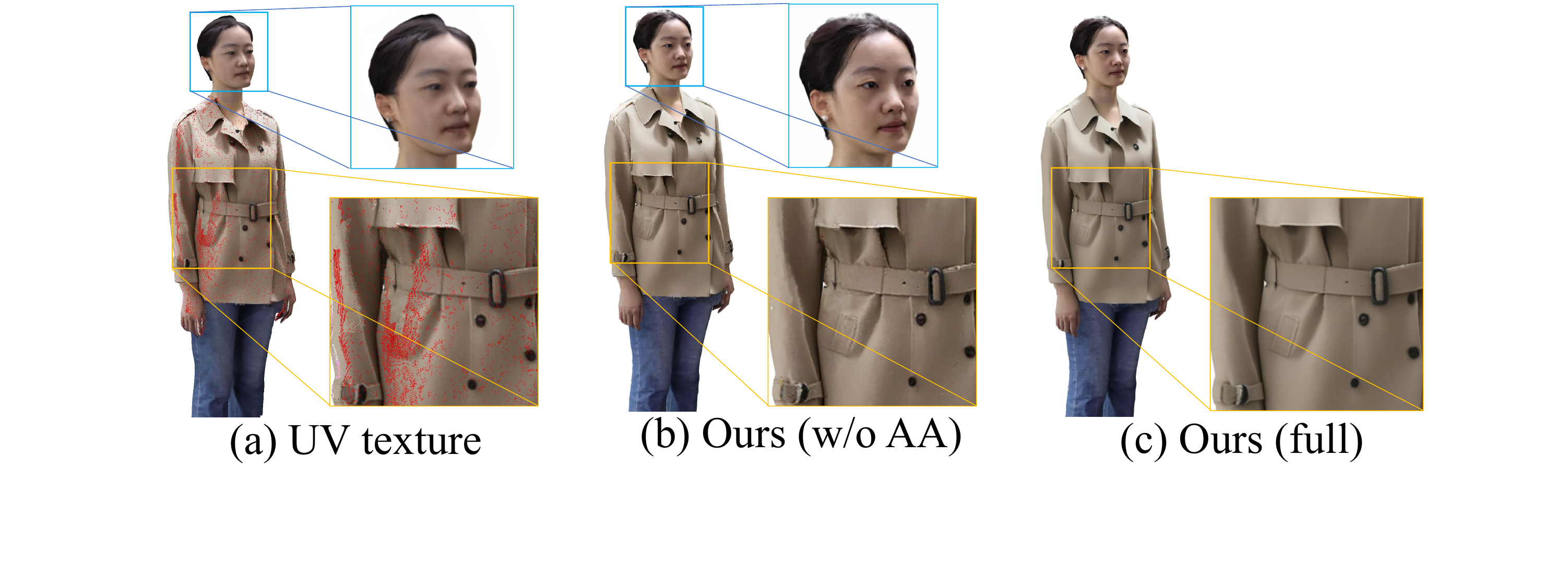}
\caption{Ablation study on the rendering capabilities of TetGS and the attribute activation stage. Zoom in for a better inspection.}

\label{fig:texture}
\end{center}
\vspace{-0.8cm}
\end{figure}

{ \noindent \bf Ablation on different appearance representations. }
Given an existing geometry, previous methods~\cite{richardson2023texture, wu2023hyperdreamer, zeng2024paint3d} paint textures using UV maps. However, due to the reconstruction error during differentiable UV rendering, they exhibit obvious artifacts in editing regions under novel views, especially for extracted meshes without predefined UV atlas, as shown in Fig.~\ref{fig:texture} (a) (red spots indicate uncolored noise).
In contrast, our method directly optimizes restricted Gaussians in 3D space using the GS rasterizer, enabling robust novel-view rendering of the editing areas, and also achieving higher reconstruction fidelity. 

{ \noindent \bf Attribute activation with augmented guidance. }
As shown in Fig.~\ref{fig:texture} (b\&c) and Tab.~\ref{tab:ablation}, the attribute activation (AA) effectively improves image quality and local details.

%% file: sec/X_suppl.tex
\clearpage
\maketitlesupplementary
\setcounter{section}{0}
\renewcommand\thesection{\Alph{section}}

In our supplementary material, we provide:
\begin{itemize}[nosep,left=1.5em]

\item Details of 3D avatar instantiation.
\item Details of localized spatial adaptation.
\item Details of texture generation.
\item Details of reference image-based editing.
\item More edited results.
\item Experimental details on baseline comparison.
\item Applications.
\item Limitations.

\end{itemize}

\section{Details of 3D avatar instantiation}
To provide accurate geometry and appearance prior for TetGS initialization and pave the way for the following editing phase, we perform high-quality 3D avatar instantiation from captured real-world monocular videos.
\subsection{Architecture of implicit reconstruction with SDF field}
\begin{figure}
\begin{center}
\setlength{\abovecaptionskip}{0.1cm}
\setlength{\belowcaptionskip}{-0.1cm}
\includegraphics[width=0.99\linewidth]{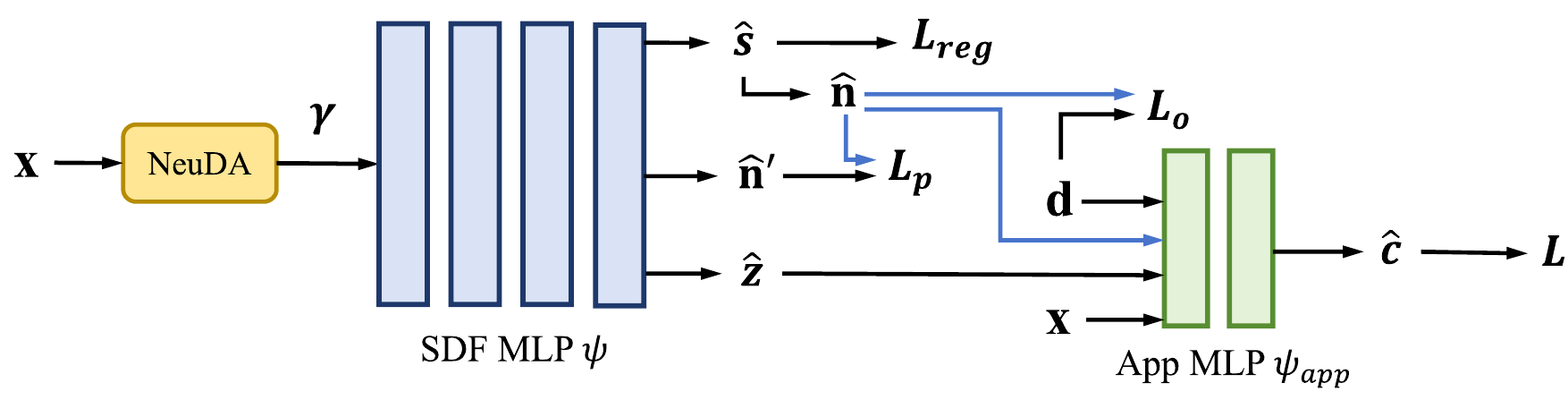}
\caption{The architecture of the implicit reconstruction with SDF field.}
\label{fig:arch}
\end{center}
\vspace{-0.5cm}
\end{figure}

To obtain precise geometric surface for TetGS initialization, we conduct multi-view surface reconstruction utilizing an SDF field, which is instantiated by a 4-layer MLP $\psi$ with 512 hidden units per layer. Given a spatial point $x$, the SDF field $\psi$ maps it to its signed distance value $\hat{s}$ to the object surface. A predicted normal $\hat{\mathbf{n}}'$ and a geometric feature $\hat{z}$ is also output by $\psi$. The SDF field is followed by an appearance field $\psi_{app}$ which predicts the view-dependent color $\hat{c}$ for point $x$ under view direction $d$. The appearance MLP $\psi_{app}$ has 2 layers with 128 hidden units. The network architecture is illustrated in Fig.~\ref{fig:arch}.
To improve sampling efficiency, we apply two rounds of proposal sampling and then a NeRF sampling following Mip-NeRF 360~\cite{barron2022mip}.
The overall training loss contains a color reconstruction loss $L_{c}$, an eikonal loss~\cite{gropp2020implicit} $L_{reg}$, and two normal regularization losses $L_p$ and $L_o$:
\begin{equation}
\ L_{rec} = L_{c} + \lambda_{reg} L_{reg} + \lambda_{p}L_p + \lambda_o L_o,
\end{equation}
where $\{\lambda_{reg}, \lambda_p, \lambda_o\}$ are set to $\{0.1, 10^{-6}, 10^{-3}\}$.

\subsection{TetGS initialization}
The reconstructed geometry is directly converted into tetrahedron grids, where we embed a different number of Gaussians for each tetrahedron. The number of Gaussians assigned to each tetrahedron is based on its extracted face area relative to the average size: faces larger than average are assigned three Gaussians, while smaller faces receive one Gaussian. The optimization of the embedded Gaussians $\mathcal{G}$ follows the original 3DGS method~\cite{kerbl20233d}, where we apply the pixel-wise reconstruction loss between the multi-view renderings $\hat{I}$ and the training images $I_{gt}$ sampled from the input monocular video:
\begin{equation}
\ L = L_1(\hat{I}, I_{gt}) + \lambda L_{SSIM}(\hat{I}, I_{gt}).
\end{equation}
To better capture high-frequency geometry and texture details, we perform TetGS initialization inside the subdivided tetrahedron grids~\cite{shen2021deep}.

\begin{table}
\setlength{\abovecaptionskip}{0.2cm}
\setlength{\belowcaptionskip}{0.0cm}
  \centering
    \caption{Quantitative evaluation (test-set view) of our method compared to 3DGS averaged on our collected dataset. 7K and 30K denote training iterations.}
      \resizebox{0.7\linewidth}{!}{
  \begin{tabular}{cccccccc}
    \toprule
    Method\big|Metric  & PSNR $\uparrow$ & SSIM $\uparrow$ & LPIPS $\downarrow$ \\
    \midrule
    3DGS-7K  & 26.12 & 0.933 & 0.195  \\
    3DGS-30K  & \textbf{27.31} & 0.941  & 0.175 \\
    Ours-7K  & 26.67 & \textbf{0.947} & \textbf{0.157}  \\
    \bottomrule
  \end{tabular}}
  \label{tab:recon}

\end{table}

\begin{figure}
\vspace{-0.0cm}
\begin{center}
\setlength{\abovecaptionskip}{0.1cm}
\setlength{\belowcaptionskip}{0.0cm}
\includegraphics[width=0.9\linewidth]{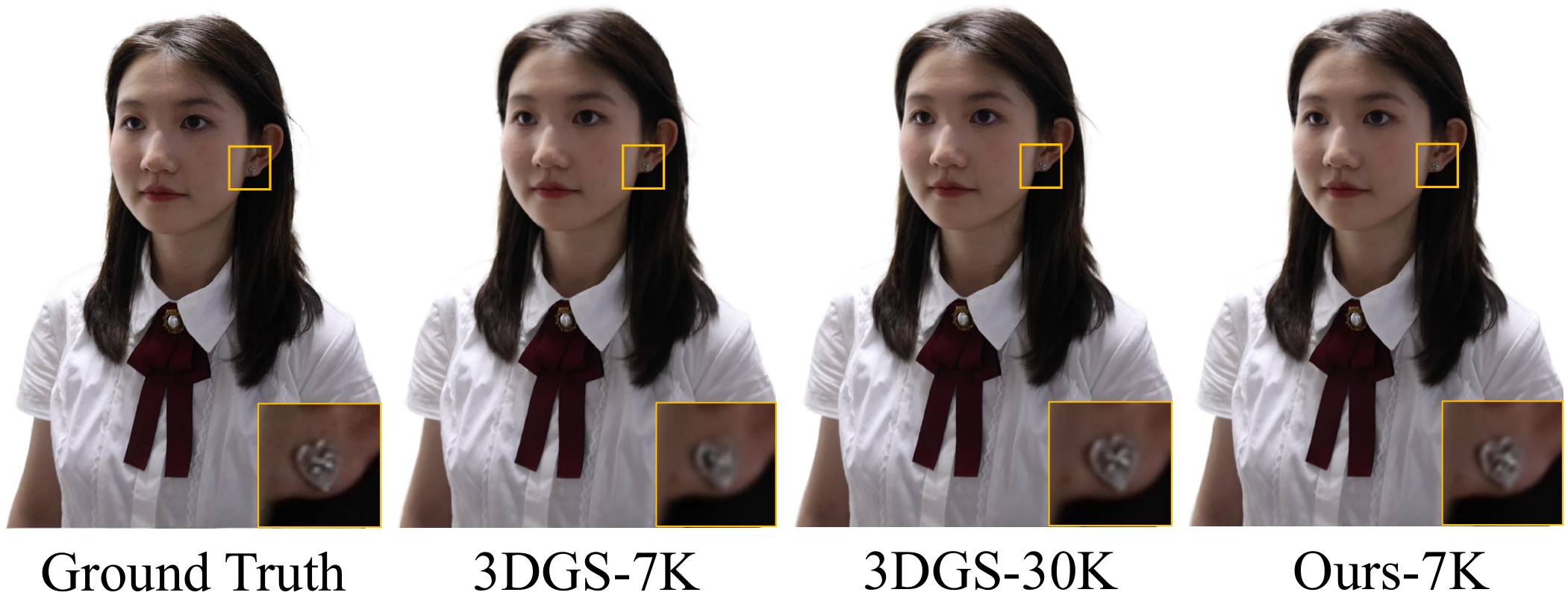}
\caption{Qualitative reconstruction comparison of TetGS against 3DGS.}

\label{fig:reviewer-1}
\end{center}
\vspace{-0.8cm}
\end{figure}

\begin{figure*}
\begin{center}
\setlength{\abovecaptionskip}{0.1cm}
\setlength{\belowcaptionskip}{-0.7cm}
\includegraphics[width=0.99\linewidth]{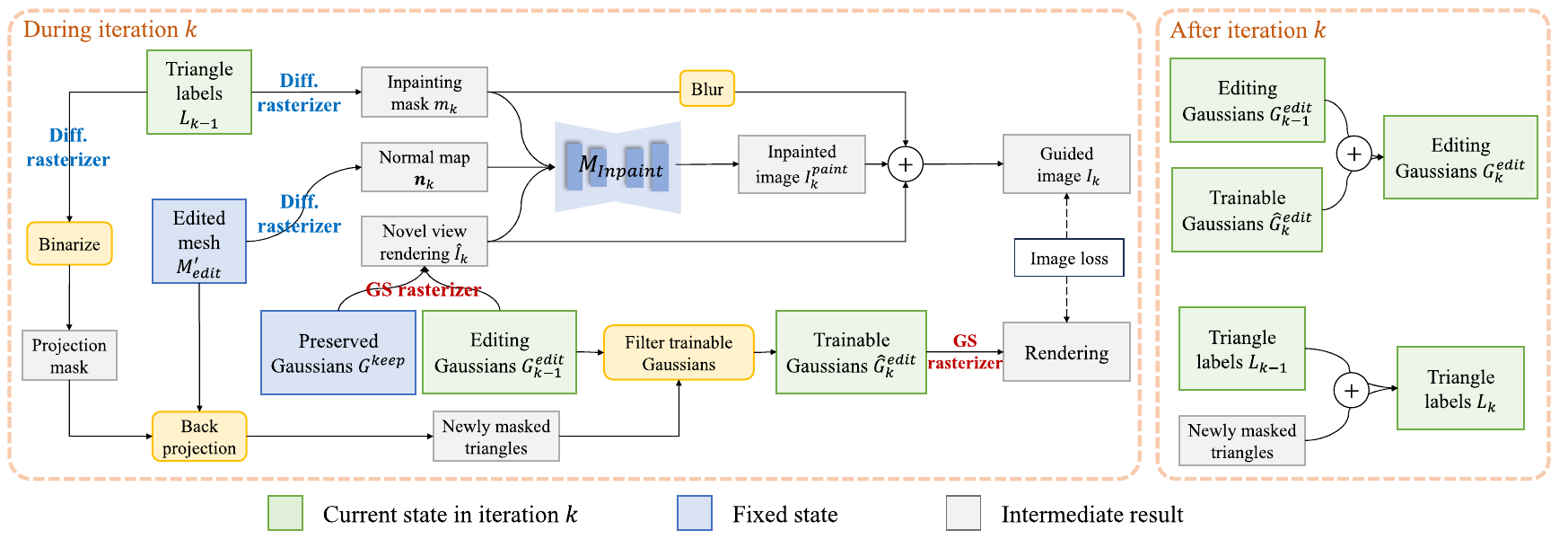}
\caption{An overview of the coarse texture generation stage.}

\label{fig:pipeline}
\end{center}
\vspace{-0.0cm}
\end{figure*}

\subsection{Reconstruction performance of TetGS against 3DGS}
Comparison of TetGS with 3DGS is shown in Fig.~\ref{fig:reviewer-1} and Tab.~\ref{tab:recon}. TetGS demonstrates comparable reconstruction performance to 3DGS, while achieving faster convergence of Gaussian parameters, benefiting from the guidance of tetrahedral grids on 3D Gaussians' spatial allocation.

\section{Details of localized spatial adaptation}
During the localized spatial adaptation of TetGS, we adopt a view sampling strategy similar to~\cite{huang2024humannorm}, focusing on both global and local regions to calculate the dual spatial constraints. The resolution of the rendered normals is $512 \times 512$. We set the global and local text prompts $y^G$ and $y^L$ as \textit{"photo of a man/woman wearing a ... garment"} and \textit{"photo of a ... garment"}, respectively.
For calculating the geometric guidance $L^G_{SDS}$ and $L^L_{SDS}$, we use a publicly available normal-adapted Stable Diffusion V1.5 model~\cite{huang2024humannorm} to  generate more detailed geometry. We set the guidance scale at 50, and anneal the timestep $t$ from $t \sim \mathcal{U}(0.02,0.80)$ into $t \sim \mathcal{U}(0.02,0.20)$. Inspired by~\cite{huang2024tech}, during the early phase of the spatial adaptation, we render normals on the coarse tetrahedron grids with a resolution of $512^3$ to encourage fast and large-scale deformation, and optimize detailed geometry within subdivided high-resolution tetrahedrons in later steps of training.

\section{Details of texture generation}
With the reallocated editing Gaussians $\mathcal{G}^{edit}$ with already learned spatial distribution, we propose to generate texture within the editing regions in a coarse-to-fine manner. The optimizer and training hyperparameters for $\mathcal{G}^{edit}$ are shared with the previous TetGS initialization stage. The guided text prompts used for texture generation are shared with the global prompt $y^G$ during the spatial adaptation stage.

\subsection{Overview of coarse texture generation}
We show an overview of the coarse texture generation stage in Fig.~\ref{fig:pipeline}, where we iteratively optimize trainable editing Gaussians under the supervision of the few-shot inpainted images. Detailed descriptions are included in Sec.~\ref{sec:tex} of the main paper.

\subsection{Diffusion guidance during texture generation}
We use the publicly available SDXL-based ControlNetPlus~\cite{controlnetplus} provided on Hugging Face for both normal-based coarse texture inpainting and I2I augmentation, which is a unified all-in-one ControlNet for image generation and editing. We integrate the normal branch and inpainting branch for the normal-based inpainter to generate consistent texture faithful to the underlying geometry, and apply the tile super resolution branch for I2I augmentation that boosts local details while preserving the original contents.
We generate the inpainted and augmented images focusing at the editing region with a resolution of $1024 \times 1024$.

\section{Details of reference image-based editing}
The proposed controllable TetGS and decoupled editing strategy naturally support reference image-guided 3D virtual try-on. We collect diverse types of garments from e-commerce websites as the reference images. The corresponding text prompts are generated by GPT-4~\cite{achiam2023gpt} with the command \textit{"describe the color and style of the garment"}.
We generate front and back-view try-on images $I_f$ and $I_b$ separately using IDM-VTON~\cite{choi2024improving}, where we input the reconstructed front or back rendering, the editing mask, the reference garment, and its corresponding text description. Since the individual and reference clothing remain consistent for both views, the generated images inherently maintain a globally coherent style. Specifically, for the back view, we add \textit{"backview"} to the text prompt to enhance image quality. The generated $I_f$ and $I_b$ serve as the guidance images for the appearance learning of the editing Gaussians, which facilitate direct texture transfer of the specific garment styles.

We also apply additional geometric supervision $L_{vton}$ during the localized spatial adaptation stage to recover faithful geometric design (Eq.10 in the main paper). The loss weights $\{\lambda_{norm}, \lambda_{mask}\}$ are set to $\{0.03, 1.0\}$.

\section{More edited results}
\begin{figure*}
\begin{center}
\setlength{\abovecaptionskip}{-0.0cm}
\setlength{\belowcaptionskip}{-0.1cm}
\includegraphics[width=0.87\linewidth]{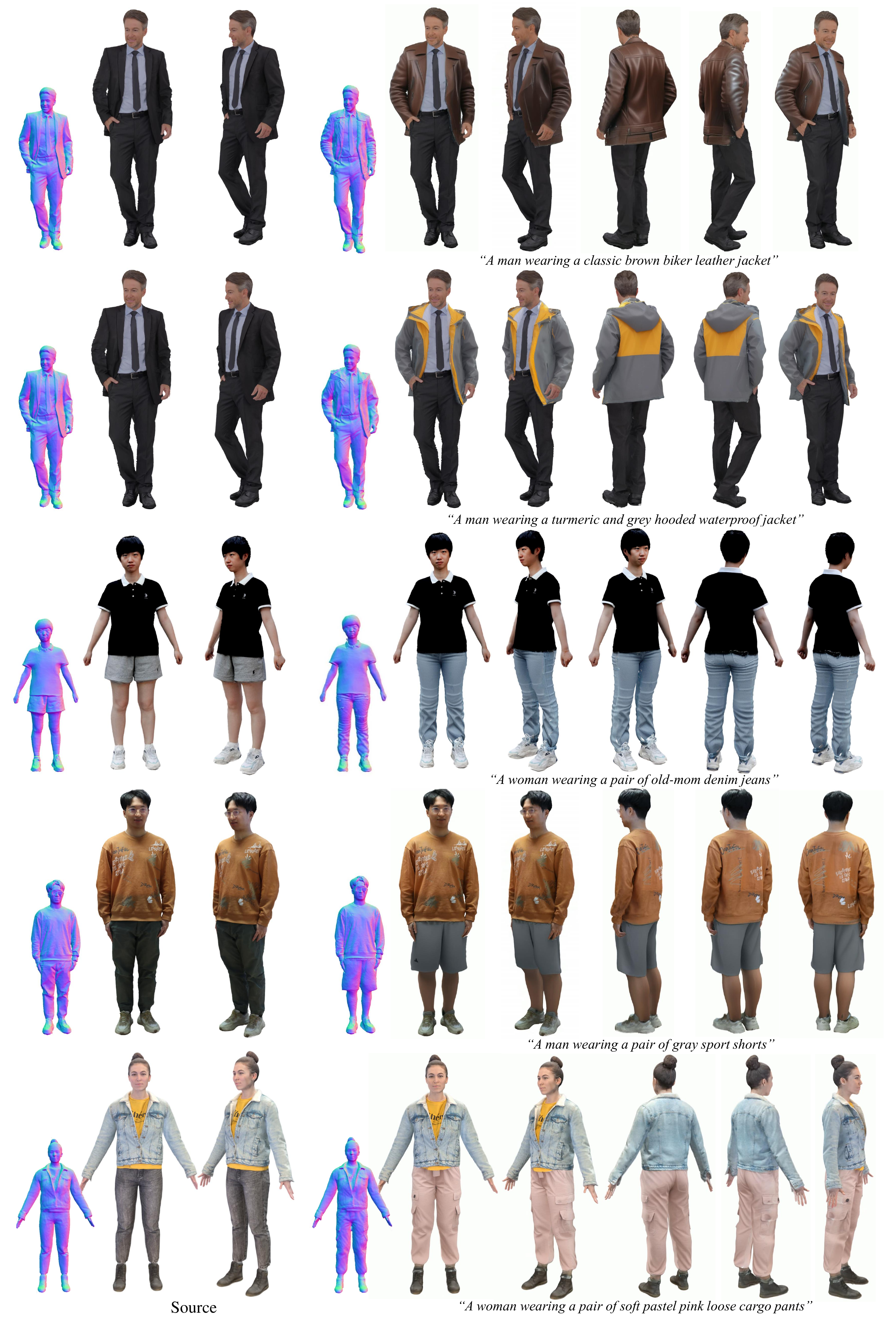}
\caption{More results on text-guided 3D avatar editing.}

\label{fig:res1}
\end{center}
\vspace{-0.5cm}
\end{figure*}

\begin{figure*}
\begin{center}
\setlength{\abovecaptionskip}{-0.0cm}
\setlength{\belowcaptionskip}{-0.1cm}
\includegraphics[width=0.84\linewidth]{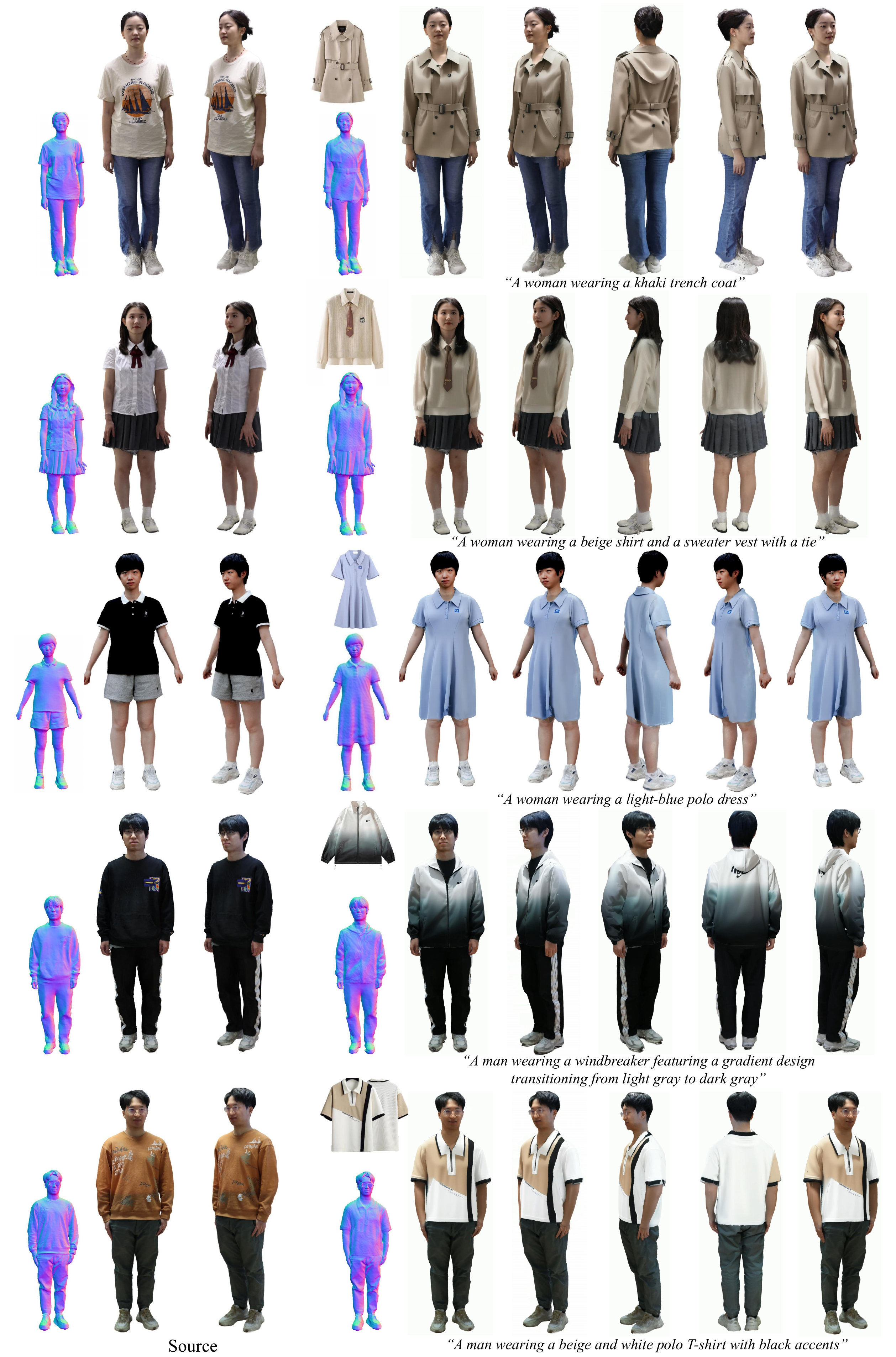}
\caption{More results on reference image-guided 3D avatar editing.}

\label{fig:res2}
\end{center}
\vspace{-0.5cm}
\end{figure*}

We showcase more edited results in Fig.~\ref{fig:res1} and Fig.~\ref{fig:res2}, including both text-guided 3D editing and reference image-based 3D virtual try-on. We demonstrate that our proposed method can handle various editing scenarios, covering upper garments, lower garments, and dresses. Our generated editable 3D avatars exhibit accurate region localization, flexible geometric adaptation, and coherent renderings with high fidelity and photorealism comparable to real-world individuals.

\section{Experimental details}
We compare our method with three 3D editing baselines: GaussianEditor~\cite{wang2024gaussianeditor}, DGE~\cite{chen2024dge} and TIP-Editor~\cite{zhuang2024tip}, which are state-of-the-art methods for text or image-guided 3D scene editing. The selected baselines cover various approaches for 3D editing with Gaussian splatting, including supervisions based on the iN2N~\cite{haque2023instruct}, the SDS loss~\cite{poole2022dreamfusion}, and multi-view consistent edited images.

{ \noindent \bf GaussianEditor. } GaussianEditor pioneers in 3D scene editing with Gaussian Splatting. It facilitates two editing models, which utilize delta denoising score~\cite{hertz2023delta} (GaussianEditor-DDS) and Instruct-NeRF2NeRF~\cite{haque2023instruct} (GaussianEditor-iN2N) as the generative guidance for editing, respectively. Since GaussianEditor-iN2N exhibits better editing performance across various scenarios, we compare our method with GaussianEditor-iN2N. To meet the instruction requirement of iN2N, we convert our text prompts into the format of \textit{"put him/her into ..."} or \textit{"give him/her a ..."}. To specify the local editing region with text-guided SAM~\cite{kirillov2023segment}, we manually provide segmentation prompts describing the interested area for GaussianEditor's semantic tracing process, which is the same prompt that we use in our method to segment multi-view masks for tetrahedron partitioning.

{ \noindent \bf DGE. } DGE is a representative 3DGS editing method that uses a multi-view consistent 2D image editor for a more stable generative supervision with spatial consistency. It utilizes Instruct-pix2pix~\cite{brooks2023instructpix2pix} as the underlying image editor and adopts spatiotemporal attention for view-consistent editing following video editing methods. DGE is built on the implementation of GaussianEditor, which also enables local semantic tracing. Thus, the editing text prompts and segmentation prompts of DGE can be shared with GaussianEditor.

{ \noindent \bf TIP-Editor. } TIP-Editor enables text-and-image-guided 3DGS editing under the supervision of the SDS loss~\cite{poole2022dreamfusion} propagated by a personalized T2I model, where a DreamBooth model~\cite{ruiz2023dreambooth} is used for original scene personalization and a LoRA layer~\cite{hu2021lora} is added for editing content personalization. We use the collected reference garment images for the LoRA training. To adapt to the concept-driven models used in TIP-Editor, we convert our editing text prompts to meet the format of \textit{"a $V_1$ man/woman wearing a $V_2$ garment"}. For fair comparison on the localized editing task, we manually set its editing bounding box close to the editing region localized by our method.

\begin{figure}
\begin{center}
\setlength{\abovecaptionskip}{0.1cm}
\setlength{\belowcaptionskip}{-0.3cm}
\includegraphics[width=0.9\linewidth]{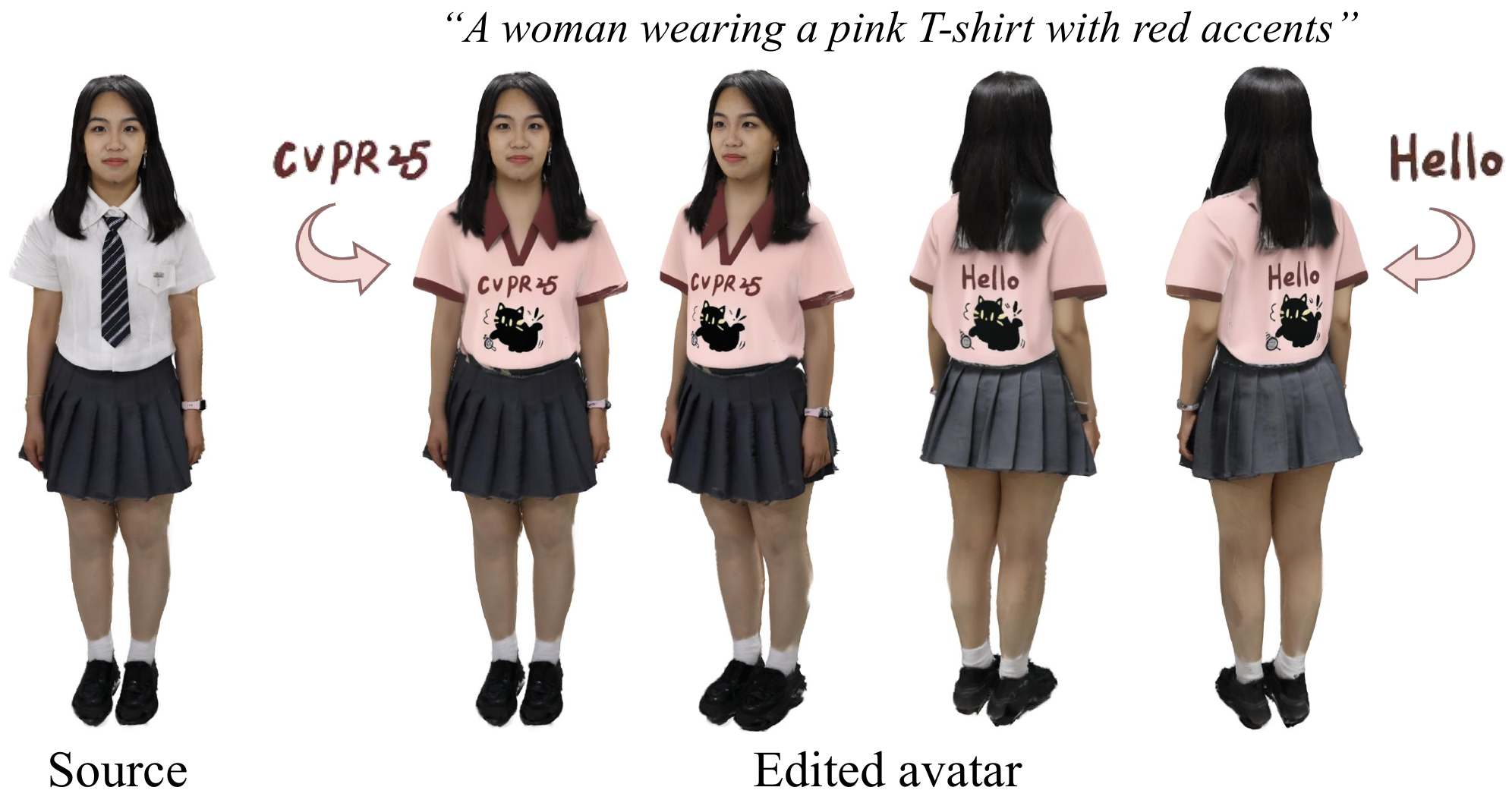}
\caption{Application on customized texture doodling.}

\label{fig:doodle}
\end{center}
\vspace{-0.2cm}
\end{figure}

\begin{figure}
\begin{center}
\setlength{\abovecaptionskip}{0.1cm}
\setlength{\belowcaptionskip}{-0.1cm}
\includegraphics[width=0.99\linewidth]{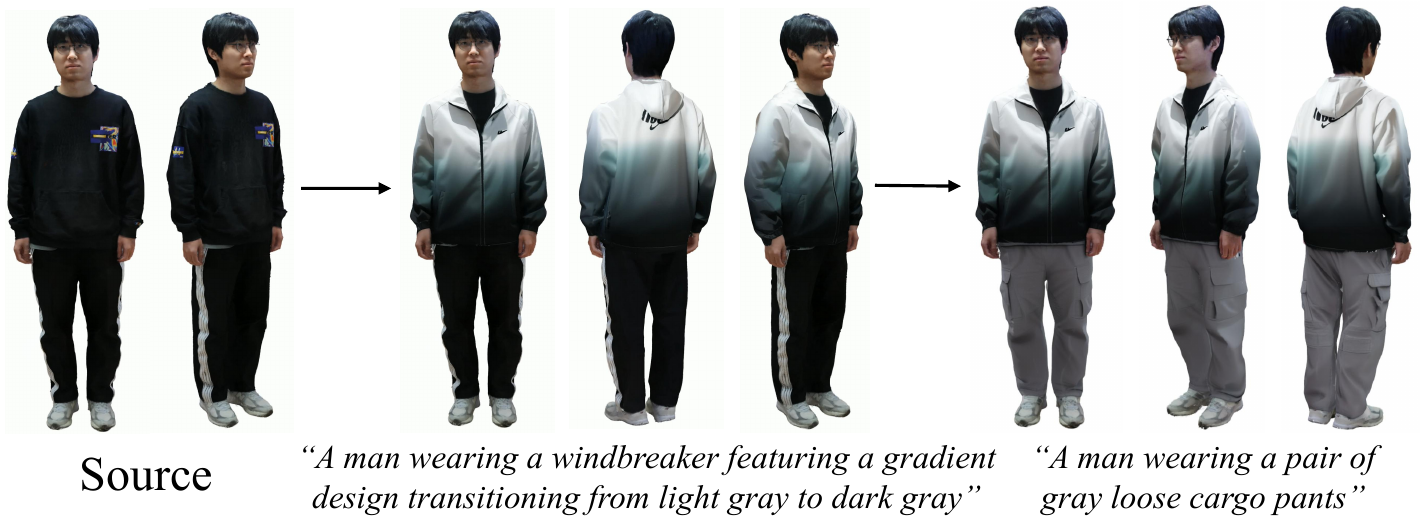}
\caption{Application on continuous editing.}

\label{fig:continuous}
\end{center}
\vspace{-0.6cm}
\end{figure}

\section{Applications}
{ \noindent \bf Texture doodling. } Benefit from the controllable TetGS representation which naturally supports decoupled geometry and appearance editing, we achieve customized texture doodling by manually editing the front and back guidance images as the supervision of the texture generation stage. Users can paint any pattern to the guidance images, and the modified texture can be directly transferred into the edited 3D avatar by optimizing Gaussian appearance under the supervision of the painted images.
We show an example in Fig.~\ref{fig:doodle}, where we paint \textit{'CVPR25'} and \textit{'Hello'} on the front and back views of the woman's T-shirt.

{ \noindent \bf Continuous editing. } Our method can continuously edit the source avatar, benefiting from our localized editing strategy. Fig.~\ref{fig:continuous} shows results of changing the upper garment followed by the pants.

\section{Limitations}

{ \noindent \bf Static human scene. }
Our method is proposed for static human scenes and the performers should stay still during the video capture. Dynamic portraits and obvious jittering may bring confused surfaces and blurred textures, due to the misalignment between multi-view observations on the pixel level.

{ \noindent \bf Extreme editing case. }
Our method may struggle to generate proper geometric changes when editing from loose garments (e.g., dresses) to tight garments, as the pose and shape of the individual's inner body are ambiguous in those situations. Adding estimated inner body prior during editing can be a potential solution to mitigate this issue.

